%% file: main.tex
\documentclass[english]{article}
\usepackage[utf8]{inputenc}
\usepackage[T1]{fontenc}
\usepackage{babel}
\usepackage{amsmath}
\usepackage{graphicx}
\usepackage{fancyhdr}

\usepackage[dvipsnames,table]{xcolor}
\usepackage[colorlinks]{hyperref}
\usepackage{multirow}
\usepackage{threeparttable}
\usepackage[margin=1.5in]{geometry}

\begin{document}

\title{Dataset of Solution-based Inorganic Materials Synthesis Recipes Extracted from the Scientific Literature}

\author{Zheren Wang\textsuperscript{1,2},
Olga Kononova\textsuperscript{1,2}, 
Kevin Cruse\textsuperscript{1,2}
Tanjin He\textsuperscript{1,2}, \\
Haoyan Huo\textsuperscript{1,2}, 
Yuxing Fei\textsuperscript{1,2},
Yan Zeng\textsuperscript{2},
Yingzhi Sun\textsuperscript{1,2},\\
Zijian Cai\textsuperscript{1,2},
Wenhao Sun\textsuperscript{3}, 
Gerbrand Ceder\textsuperscript{1,2{*}}}
\date{}
\maketitle
\thispagestyle{fancy}

\noindent\textsuperscript{1}Department of Materials Science and Engineering, University of California, Berkeley, CA 94720, USA; \\
\textsuperscript{2}Materials Sciences Division, Lawrence Berkeley National Laboratory, Berkeley, CA 94720, USA; \\
{\textsuperscript{3}Department of Materials Science and Engineering, University of Michigan, Ann Arbor, MI, USA }\\
{*}Corresponding author: Gerbrand Ceder (gceder@berkeley.edu)\\

\newpage
\begin{abstract}

    The development of a materials synthesis route is usually based on heuristics and experience.
    A possible new approach would be to apply data-driven approaches to learn the patterns of synthesis from past experience and use them to predict the syntheses of novel materials.
    However, this route is impeded by the lack of a large-scale database of synthesis formulations. 
    In this work, we applied advanced machine learning and natural language processing techniques to construct a dataset of 35,675 solution-based synthesis ``recipes'' extracted from the scientific literature. 
    Each recipe contains essential synthesis information including the precursors and target materials, their quantities, and the synthesis actions and corresponding attributes. 
    Every recipe is also augmented with the reaction formula. 
    Through this work, we are making freely available the first large dataset of solution-based inorganic materials synthesis recipes.
    
\end{abstract}

\section{Background \& Summary}

    Big-data-driven approaches have helped to establish a new paradigm of scientific research\cite{CSforSci, Paradigm4, FischerNatMat}. 
    In materials science specifically, the Materials Genome Initiative (MGI) effort has significantly facilitated and accelerated materials discovery and design by deploying large-scale \textit{ab initio} computation and building computed databases of structure--property relationships \cite{MGIreport, pymatgen, MatProj}.
    Unlike computational data, the experimentally determined properties and structures of inorganic materials are mainly available in manually curated databases\cite{ICSDoriginal, ICSDnew, NISTwebbook, PaulingFile, PCD}.
    Such well-curated experimental databases have led to early machine learning models that address difficult problems in materials research, such as structure prediction\cite{FischerNatMat}.
    The ability to efficiently design and predict the structure of novel advanced materials with the assistance of computed and experimental databases has shifted the materials innovation challenge toward understanding and determining the synthesis routes for novel materials\cite{SumpterCNPGM}. 
    In principle, this challenge could also be mastered using data-driven approaches, since very few guiding principles are available for materials synthesis. \cite{BianchiniNatMat, SunSciAdv}
    Indeed, in organic chemistry, AI-guided synthesis planning \cite{DuvenaudNIPS,LeyReview} has already been successfully implemented in certain cases, such as in predicting retrosynthesis \cite{SeglerNature} and in complex natural product synthesis design \cite{CompPlanningNature}.
    Although datasets for the synthesis of organic materials are widely available \cite{Reaxys, Pubchem}, there is not yet a large-scale database of inorganic synthesis routes, which is needed to train advanced deep-learning models to enable a breakthrough in AI-assisted design and optimization of inorganic materials synthesis.

    Scientific publications represent the largest repository of knowledge about material synthesis and can be used as a reliable source of data. 
    However, human-written descriptions of syntheses require additional levels of interpretation for conversion into a codified, machine-operable format. 
    Additionally, manual extraction of synthesis information is laborious, even for a very limited number of papers\cite{kononovaISCIENCE, Gaultois2013, Ghadbeigi, oh2016meta}.
    Given these obstacles, an automated information extraction pipeline can accelerate data collection and assist in building structured synthesis ``recipes'' from scientific text. 
    Natural language processing (NLP) approaches have been widely developed in the past decade, and various advanced tools for high-quality information extraction from unstructured text are available to researchers.

    In materials science, NLP has been used to extract and analyze materials properties \cite{ColeData, court2020magnetic, CDEbattery}, applications \cite{onishi2018relation,tshitoyan2019unsupervised}, and synthesis conditions for some limited cases \cite{OlivettiCM}. 
    Various NLP tools, including ChemDataExtractor \cite{ChemDataExtractor}, OSCAR\cite{OSCAR4}, ChemicalTagger\cite{ChemicalTagger}, and others \cite{IRreview, ChNERrev, kononovaISCIENCE}, have been developed to extract information from chemical text.
    Recently, advanced models based on deep convolutional and recurrent neural networks \cite{Korvigo2018a, he2020similarity, westonJCIM2019, kuniyoshi2020annotating} have been proposed to improve the accuracy of chemical data extraction.

    Text-mining approaches in materials science have also been used to construct automated pipelines for collecting information about materials synthesis from publications and to build large-scale publicly available datasets from such collected data, including datasets of synthesis formulations for metal oxides\cite{OlivettiData, OlivettiCM, Kim2017d}, germanium-containing zeolites\cite{jensen2019machine}, and perovskites\cite{Kim2020}. 
    In recent work, our group has developed a text-mining pipeline to construct the first large-scale dataset of solid-state ceramics synthesis ``recipes'', which includes not only the starting materials and final products but also the synthesis actions, their attributes, and balanced chemical-reaction equations \cite{SolidStateDataPaper}.

    In the current work, we built a more advanced extraction pipeline (Figure.~\ref{fig:fig1})  which uses various advanced machine learning and natural language processing techniques to extract precise recipe data for solution-based inorganic materials synthesis procedures from the scientific literature.  
    Solution recipes are considerably more complex than solid state synthesis and require the precise extraction of not only the chemicals involved but also their respective amounts (since they determine concentration in solution).  
    In addition, more complex organic and mixed organic-inorganic compounds are used to solubilize ions or to control solution conditions.
    By applying the extraction pipeline, we codified 35,675 solution-based inorganic materials synthesis “recipes” from over 4 million papers.
    Extracted information includes target material and precursors, their quantities, and the synthesis operations and their attributes. 
    Information about the targets and precursors is then used to build a reaction formula for every synthesis procedure. 
    This dataset is the first large-scale dataset of solution-based synthesis recipes, and should pave the way for future data-driven approaches to inorganic materials synthesis and synthesizability, and to design optimized synthesis procedures in automated experimentation.

\section{Methods}

\subsection{Content acquisition}

    The journal articles used in this work were downloaded with publisher consent from Wiley, Elsevier, the Royal Society of Chemistry, the Electrochemical Society, the American Chemical Society, the American Physical Society, the American Institute of Physics, and Nature Publishing Group.
    A customized web-scraper, Borges (see Codes Availability section below), was used to automatically download a broad selection of materials-relevant papers published after the year 2000 from publishers' websites in HTML/XML format.
    We selected 2000 as the cutoff year as parsing of materials science papers stored as image PDFs (as for most papers published before 2000) introduces a significant number of errors due to the limitations of currently available optical character recognition models on chemistry containing text\cite{mouchere2016advancing, mahdavi2019icdar}.
    
    To convert the articles from HTML/XML into raw-text files, we developed the LimeSoup toolkit (see Codes Availability section below), which takes into account the specific format standards of various publishers and journals.
    The full-text and metadata of the articles such as the journal name, article title, abstract, author names, etc., are stored in a MongoDB (\url{www.mongodb.com}) database collection.
    To date, we have accumulated 4.06 million articles, which are used for further processing down the pipeline (Figure~\ref{fig:fig1}).

\subsection{Paragraph classification} \label{sec:paragraphs_classification}

    Paragraphs containing information about solution synthesis (referred to as ``synthesis paragraphs'' throughout this paper) were identified using a Bidirectional Encoder Representations from Transformers (BERT) model \cite{devlin2018bert}.
    The model was pre-trained on full-text paragraphs of 2 million papers randomly drawn from our database in a self-supervised way, i.e., by predicting masked words based on their surrounding context.
    After training the BERT model, we fine-tuned the paragraph classifier using 7,292 paragraphs labeled as either ``solid-state synthesis'', ``sol-gel precursor synthesis'', ``hydrothermal synthesis'', ``precipitation synthesis'', or ``none of the above''.
    The resulting F1 score of the paragraph classification is 99.5\%, an improvement over the F1 score of 94.6\% in our previous work \cite{HuoCNPGM}, when evaluated using the same labeled training dataset.

\subsection{Synthesis recipe extraction}
    
    A solution-based synthesis recipe includes the precursors and target materials, their quantities, and the synthesis actions and their attributes, properly sequenced.
    This is the minimum essential information required to complete a synthesis route.
    A schematic representation of the recipe is shown in the bottom panel in Figure~\ref{fig:fig1}.
    In the sections below, we provide a brief overview of the methods used for each step of the recipe extraction.

\subsubsection{Materials entity recognition (MER)} \label{sec:MER}

    Materials entities in synthesis paragraphs are identified and classified as \emph{target}, \emph{precursor}, or \emph{other} via a two-step sequence-to-sequence model as introduced in our previous work \cite{he2020similarity}.
    In the current work, we replaced the original Word2Vec embedding model used previously \cite{NIPS2013_5021} with a BERT model trained on papers from the materials science domain (see Section \ref{sec:paragraphs_classification} above).
    First, each word token was transformed into a digitized BERT embedding vector.
    A bi-directional long-short-term memory neural network with a conditional random-field top layer (BiLSTM-CRF) was used to determine whether the token was a materials entity or a regular word, and each materials entity was replaced with the keyword \texttt{<MAT>} before being classified as either a \emph{target}, \emph{precursor}, or \emph{other} material using a second BERT-based BiLSTM-CRF network. 
    In addition to the 834 annotated solid-state synthesis paragraphs from 750 papers used in our previous work\cite{he2020similarity}, we manually annotated 447 solution-based synthesis paragraphs from 405 papers by labeling each word token as \emph{material}, \emph{target}, \emph{precursor}, or \emph{outside}.
    The annotated dataset was split into training, validation, and test sets with a paper-wise ratio of 700:150:305 to train the aforementioned two neural networks.

\subsubsection{Extraction of synthesis actions and attributes} \label{sec:Operation}

    We implemented an algorithm which combines a neural network and sentence dependency tree analysis to identify synthesis actions in the text.
    First, the Word2Vec model from the Gensim library \cite{rehurek_lrec} was re-trained on $\sim$400,000 synthesis paragraphs of four synthesis types (see Section~\ref{sec:paragraphs_classification} above).
    These word embeddings were used as the input for a recurrent neural network that takes a sentence word-by-word and assigns labels to the verb tokens: \emph{not-operation}, \emph{mixing}, \emph{heating}, \emph{cooling}, \emph{shaping}, \emph{drying}, or \emph{purifying}. 
    For each obtained synthesis action, we parsed a dependency sub-tree using the SpaCy library \cite{SpaCy} to obtain information about the corresponding temperature, time, and environment.
    To extract the corresponding values of these attributes, we used a rule-based regular expression approach \cite{jurafsky2009speech}.

\subsubsection{Extraction of material quantities}

    To extract the numerical values of material quantities and assign them to the corresponding materials obtained using the MER model (see Section~\ref{sec:MER} above), we applied a rule-based approach to search along the syntax tree \cite{jurafsky2009speech}.
    The NLTK library \cite{NLTK} was used to build the syntax trees for each sentence in a paragraph. 
    The words in given sentences are leaf nodes of syntax trees. 
    We then applied an algorithm to cut the syntax tree of each sentence into the largest sub-trees for every material, with each sub-tree having only one material entity: 1. we first identified the materials on leaf nodes; 
    2. starting from each material, we identified the largest sub-trees, i.e., we traversed the syntax tree upwards until there was more than one material leaf node descending from the same node;
    3. the largest sub-tree for a given material was defined as the sub-tree formed by the node and its descendants identified in step 2.
    Next, we searched for the quantities in each sub-tree given as molarity, concentration, or volume. 
    Finally, we assigned the quantities found to the unique material entity in the sub-tree.

\subsubsection{Building reaction formulas}

    For every synthesis procedure described in a paragraph, we built a chemical formula.
    Every material entity was converted from a text-string representation into a chemical-data structure using an in-house material parser toolkit (see Codes Availability section below). 
    The data structure included information about the material formula, composition, and ions. 
    We then paired the target with precursors containing at least one element in the target except for hydrogen and oxygen and defined those precursors as ``precursor candidates''.
    Next, we computed the oxidation state change of elements from each ``precursor candidate'' to the target and determined whether the precursor was oxidized or reduced. 
    If precursors were reduced or oxidized, we also included the corresponding redox agents in the reaction formula.
    The agents can either be another ``precursor candidate'' or a commonly used oxidizing or reducing agent from the remaining material entities marked as \emph{other} or \emph{precursor} by the MER algorithm (see Section~\ref{sec:MER} above).

\subsection{Dataset generation}

    The dataset generation followed the protocol displayed in Figure~\ref{fig:fig1}.
    We downloaded a total of 4,061,814 papers using web scraping and identified the experimental sections by keyword matching in section headings, with keywords including ``experiment'', ``synthesis'', ``preparation'', and their morphological derivations. 
    ChemDataExtractor \cite{ChemDataExtractor} was used to split the plain-text paragraphs into sentences and words. 
    After classification (see Section \ref{sec:paragraphs_classification} above), 364,076 paragraphs describing solid-state, hydrothermal, sol-gel, and precipitation syntheses were obtained.
    Among them, 189,553 paragraphs described hydrothermal or precipitation syntheses, which we categorize as solution-based synthesis methods.
    These paragraphs were further processed to extract the precursors, targets, quantities, and operations with corresponding conditions and to build the reaction formula (Figure~\ref{fig:fig1}).

\section{Data Records}
 
    The solution-based synthesis dataset is provided as a single JSON file, available at\\ \url{https://doi.org/10.6084/m9.figshare.16583387.v1}. 
    There are 20,037 hydrothermal synthesis reactions and 15,638 precipitation synthesis reactions. 
    Each record corresponds to a synthesis recipe extracted from a paragraph and is represented as an individual JSON object.
    If a paragraph reported the synthesis of several materials, the corresponding reactions were split into separate data records.
    In addition to the chemical formula, the metadata for each reaction returns the data structure used in our previous work\cite{SolidStateDataPaper}, which includes: DOI of the paper, a snippet  of the corresponding synthesis paragraph (50 first and 50 last characters to facilitate its lookup), chemical information about the target and precursor materials used in the reaction, and operations with their corresponding attributes.
    We also included the materials with their corresponding quantities in the metadata.
    The details of the data format are given in Table \ref{tab:tab1}
    
    The chemical formula for the reaction is stored as a string (\texttt{reaction\_string}) as well as in a dictionary containing lists of precursors (\texttt{left\_side}) and target materials (\texttt{right\_side}) in \texttt{reaction}.
    
    The metadata for target materials and precursors used to construct the chemical formula are represented by the following data structure:
    \begin{itemize}
        \item \texttt{material\_string}: string of material as given in the original paragraph before being parsed into a chemical composition.
        \item \texttt{material\_formula}: chemical formula associated with the material (given originally or constructed empirically by parser).
        \item \texttt{composition}: chemical composition of the material derived from its formula.
        Aside from single-compound materials, we found that a large portion of the materials (predominantly target materials) are composites, mixtures, solid solutions, or alloys written as a sequence of compound-fraction pairs. 
        Therefore, a chemical-composition entity is represented by a list of dictionary entries, where each item is associated with a compound found in the materials formula. 
        The fraction of each compound in the material is given in \texttt{amount}, and its chemical composition (i.e., the elements and stoichiometry) is given in \texttt{elements}. 
        If a material is one compound, the list has only one item and \texttt{amount}=1.0. 
        If a material is a hydrate, water is added to the \texttt{composition} list with \texttt{amount} corresponding to the amount of water molecules (if specified).
        \item \texttt{additives}: list of additive elements (i.e., elements used for doping, stabilization, or substitution) resolved from the material string.
        \item \texttt{elements\_vars}: lists all variable elements and their corresponding values found in the materials. 
        \item \texttt{amounts\_vars}: lists all variable element ratios and their corresponding values found in the material formula. 
        The values of each variable are given as a structure with \texttt{values} listing the values of each specific variable and \texttt{max\_value}/\texttt{min\_value} values if a range is given in the paragraph.
        \item \texttt{oxygen\_deficiency}: yes/no attribute that reflects whether a material was synthesized with unspecified oxygen stoichiometry. 
        \item \texttt{mp\_id}: ID of the lowest-energy polymorph entry in the Materials Project database (\href{https://materialsproject.org/}{materialsproject.org}) if the material is found there. 
    \end{itemize}
    
    To facilitate querying of the dataset, the \texttt{targets\_string} field contains the target material formulas, and the \texttt{solvents} field contains all solvent(s) from matching material entities marked as \emph{other} by the MER model with a table of common solvents adopted from Common Solvents Used in Organic Chemistry (\href{https://organicchemistrydata.org/solvents/}{organicchemistrydata.org/solvents/}).

\section{Technical Validation}

\subsection{Extraction completeness and accuracy}

    To ensure high accuracy of the dataset, we included only those data that produced complete reaction formulas at the final step of the pipeline.
    This strategy reduced potential errors in the dataset that may have been caused by composition-parsing failure, incomplete extraction, or incomplete information provided by the text. 
    We applied the extraction pipeline to 189,553 solution-based synthesis paragraphs, 28,749 of which generated a reaction formula, giving an extraction yield of $\sim$15\%. 
    To evaluate the source of the loss, we randomly selected and manually checked 100 solution-based synthesis paragraphs that did not produce any reactions. 
    Among those 100 paragraphs, 36 were written with an incomplete list of precursors or targets in the text, such that human experts would not be able to reconstruct the reaction based solely on the information provided in the paragraph. 
    For the remaining 64 paragraphs, the loss was due to: 1. the use of organic precursors with complex groups or complicated notation (e.g., acronyms) that could not be parsed into a chemical composition by our parser or 2. MER misidentification resulting in an incomplete or incorrect list of precursors and (or) target entities such that the reactions could not be built.

    To evaluate the quality of the dataset, we had a human expert test 100 randomly pulled entries.
    The human expert manually extracted the information presented in the recipe, and the results were compared with those extracted by the pipeline. 
    Table ~\ref{tab:tab2} presents the accuracy statistics, which include the precision, recall, and F1 scores calculated from the tested entries.
    For the fields that included reaction, targets, precursors, operations, operation temperatures, time, and atmosphere, the F1 scores were over 90\%. 
    The relatively low recall, and hence F1 score, for the extraction of materials quantities can be mainly explained by the MER algorithm missing the corresponding material entity and, thus, the quantities not being assigned. 
    The accuracy of the obtained dataset is comparable to that in our previous work\cite{SolidStateDataPaper} and of other text-mined datasets\cite{OlivettiData}.

\subsection{Data exploration analysis}

    To test the diversity of the dataset and its coverage of the materials space, we analyzed unique materials (targets and precursors) and reactions.
    The dataset contains 11,603 unique reactions that include 2,870 unique precursors and 5,416 unique targets.
    The ten most frequent targets in the dataset and their corresponding precursors are listed in Table~\ref{tab:tab4}. 
    The target list captures materials that have drawn substantial attention in the past two decades: catalysts (ZnO, Fe\textsubscript{2}O\textsubscript{3}, TiO\textsubscript{2}, Fe\textsubscript{3}O\textsubscript{4}, SnO\textsubscript{2}, ZrO\textsubscript{2}, CuO), adsorbents (SiO\textsubscript{2}), various materials for sensors (ZnO, Fe\textsubscript{2}O\textsubscript{3}, WO\textsubscript{3}), quantum dots (CdS), and semiconductors (ZnO, TiO\textsubscript{2}, SnO\textsubscript{2}, CdS). 
    Unsurprisingly, these most frequent target materials usually appear in multiple applications, as they possess desirable physical and chemical properties in many scientific and engineering fields. 
    
    We use the periodic table representation (Figure~\ref{fig:fig2}) to visualize the chemical space covered by the dataset. 
    For each element, the fraction of synthesis recipes containing this element in the target formula is shown with the yellow-to-navy blue gradient framed at the top of each element box. 
    The most data-rich elements are transition metals in the third period, such as Zn, Fe, Ti, Ni, and Co, in accordance with the compounds listed in Table~\ref{tab:tab4}. 
    The next-most prevalent targets are materials with Bi, Sn, Al, W, Mo, Cu, Zr, or Li. 
    The least common elements are rare elements such as Ru, Rh, Hf, Ta, Re, and Ir.
    The elements Fr, Ra, Tc, and Pm are not present in the target materials of the dataset.
    Additionally, we calculated the frequency of co-occurrence of chemical elements and common ions in precursor materials to understand how different ions are brought into solution.
    In Figure~\ref{fig:fig2}, the frequencies for each ion are displayed as colored bars. 
    The length of the bar is the fraction of one specific ion paired with the element normalized over all precursors for this element.

    The results displayed in the periodic table (Figure~\ref{fig:fig2}) agree well with several rules-of-thumb for the selection of precursors usually applied by scientists to synthesize materials \cite{WangNanoResearch}.
    First, the commonly used precursors are mainly those that are widely available from companies such as Sigma-Aldrich and Fisher Scientific. 
    For example, Li\textsubscript{2}CO\textsubscript{3} or LiOH for Li and sulfate or chloride for Fe. 
    Second, solubility also affects the choice of precursors. 
    For instance, inorganic salts (such as nitrates, sulfates, and chlorides) are commonly used because of their high solubility \cite{WangNanoResearch}.
    Third, the organization of the periodic table reflects relationships between the various element properties. 
    We observed that precursors with neighboring elements in the periodic table tend to have similar ions paired. 
    For instance, nitrates, sulfates, and chlorides are commonly used anions for 3rd-period transition metals, whereas the precursors for lanthanides are mostly oxides and nitrates.
    
    We used information about the extracted materials and sequences of synthesis actions to classify the solution-based synthesis recipes into four categories of synthesis protocols (table in Figure~\ref{fig:fig3}) according to the following definitions: 
    \begin{itemize}
    \item \textit{solution-mixing with firing step} has a final firing step after the precipitate is obtained from the solution;
    \item \textit{aqueous solution synthesis} has no final firing step after precipitating the compound from the solution and the solvent is water;
    \item \textit{non-aqueous solution synthesis} has no final firing step after precipitating the compound from solution and the solvent(s) is (are) organic;
    \item \textit{aqueous--non-aqueous mixed solution synthesis} has no final firing step after precipitating the compound from solution and the solvents are a mixture of water and organic solvent(s).
    \end{itemize}

    The resulting distributions of synthesis protocols over the aforementioned categories are shown in the two pie charts in the top-right corner of Figure~\ref{fig:fig3}.
    Note that as solution-based synthesis includes both hydrothermal and precipitation synthesis according to our definition (see Section~\ref{sec:paragraphs_classification}), we analyzed these synthesis types separately.
    As observed in the pie charts, only 20\% of the recipes in the hydrothermal synthesis subset have a firing step after solution mixing.
    Among those that do not have firing step, 63\% use only water as a solvent, 8\% use only organic solvents, and 9\% use both water and organic solvents. 
    In contrast, the fractions in the precipitation synthesis subset are 43\%, 46\%, 5\%, and 6\%, respectively.

    These results are as expected.
    A firing step after solution mixing plays various roles in synthesis. 
    For example, it can dehydrate the targets, decompose the intermediates to produce the final products, change the oxidation state, change the morphology, or improve crystallization. \cite{firemorphlogy,fireoxide,firedecomp}.
    Thus, the decision of whether to use firing after solution steps strongly depends on the chemical properties of the target. 
    To explore this in more detail, we split the targets according to their anion type (oxide, sulfide, etc.) and different oxidation states of several data-rich transition-metal elements.
    We then computed the distribution of synthesis categories for each of the split subsets.
    Figure~\ref{fig:fig3} presents the results for the most prevalent subsets of oxides, sulfides, and elements  Fe\textsuperscript{2+}, Fe\textsuperscript{3+}, Co\textsuperscript{3+}, Ni\textsuperscript{2+}, Cu\textsuperscript{2+}, and Zn\textsuperscript{2+}.
    The fraction of recipes with a firing step in precipitation synthesis is larger than that in hydrothermal synthesis. 
    This observation holds for all targets, all oxides, all sulfides, and individual oxides and sulfides with queried oxidation states of transition metals.  
    This finding can be also interpreted as hydrothermal synthesis often being used to obtain final products in a ``one-shot'' process, without subsequent firing after solution mixing, likely because many compounds can be crystallized as anhydrous powders with controlled size and morphology directly from hydrothermal synthesis.
    According to the standard hydrothermal synthesis procedure, the reaction is performed  under high pressure and at a temperature higher than the boiling point of the solvent under normal pressure in an autoclave.
    In contrast, precipitation synthesis is performed under normal pressure. 
    The higher temperature and high pressure applied during hydrothermal synthesis, relative to the ambient conditions typically used in precipitation methods, are associated with enhanced kinetics in chemical transport, nucleation, and crystal growth and thus with a more effective dissolution--recrystallization process. 
    Furthermore, the physico-chemical properties, such as the viscosity and dielectric constant of water or other solvents, change pronouncedly under conditions of hydrothermal synthesis, affecting the solubility and mobility of species in the solution and eventually facilitating crystallization \cite{BYRAPPA20131}.
    Therefore, hydrothermal synthesis does not need a firing step as often as precipitation synthesis.
    
    The hydrothermal synthesis data subset has more records for sulfides than the precipitation subset.
    One possible reason is that the desired morphology of sulfides can be achieved directly from hydrothermal synthesis, as we discussed before\cite{BYRAPPA20131}.
    Another possible reason might be that most sulfides are not thermally stable and may decompose under firing temperature.
    The thermal instability also explains why the fraction of recipes with a firing step for sulfides is significantly lower than that for oxides.
    Furthermore, sulfide synthesis often requires temperatures higher than ambient conditions. 
    Temperatures in precipitation synthesis are not high enough to trigger reactions to produce sulfides \cite{kulkarni2017sulfides, BERHAULT2016313}. \\

    Solution-based synthesis is an important area of materials synthesis \cite{HuoCNPGM}.
    It provides the ability to precisely control the morphology of the synthesized specimen\cite{WangNanoResearch} or synthesize metastable phases\cite{SunNC19}.
    In order to extract solution-based synthesis recipes, we have built NLP/ML approaches wrapped into an automated text-mining pipeline: 1. we annotated new solution-based synthesis experimental data; 2. we incorporated a pre-trained BERT model to predict the type of synthesis described in paragraphs as well as to recognize material entities; 3. we implemented a recurrent neural network to extract synthesis actions; 4. we built chemical formulas from material entities based on chemistry knowledge; and 5. we developed a new rule-based method to extract the quantities of the materials.

    Nevertheless, challenges remain in the mining of scientific literature and construction of robust and accurate large-scale datasets.
    First, the organic precursors with complex radicals commonly used in solution-based synthesis pose a challenge for parsing and extracting chemical information.
    Constructing reaction formulas becomes problematic when the precursor information is lost. 
    Therefore, these entries are mostly dropped out later in the pipeline.  
    To address this issue, a universal parser that can parse chemical tokens needs to be developed.

    Second, our data were extracted from the experimental section from the main body of each paper and do not include any information about the actual synthesis results, e.g., whether the material was synthesized using the reported procedure or which structure was obtained.
    This problem can be overcome by introducing a model that can parse characterization data (e.g., X-ray diffraction patterns or electron microscopy images) and relate them to the corresponding synthesis conditions.
    To the best of our knowledge, this task has not been performed to date and is the subject of future efforts.
    Even though the actual results of a synthesis can be extracted from a paper, there remains the challenge of data interpretation and usage, as the authors usually report only successful and ``cherry-picked'' experimental results. 
    This introduces significant anthropogenic bias toward ``positive'' data with little ``negative'' content in the dataset, thus limiting the tasks for future machine-learning applications \cite{Raccuglia, jia2019anthropogenic}. 
    A promising approach to solve this issue is to incorporate results obtained by autonomous robotic synthesis platforms that can provide a vast amount of ``negative'' data in a reasonable time frame \cite{Burger2019Nature, SzymanskiMatHor}.

    Finally, solution-based synthesis is advantageous when the control of specimen morphology is required, e.g., when synthesizing noble-metal nanoparticles.
    However, this dataset does not provide information about the morphology of the synthesized materials, though such information is often contained in characterization or results paragraphs instead of the experimental section.
    The extraction of morphology and other solution synthesis outcomes is another text-mining challenge in materials science research that requires the development of advanced algorithms and models \cite{kononovaISCIENCE}, which is beyond the scope of the current study.

\section{Usage Notes}
    
    The dataset is provided in JSON format as a single file. 
    All major programming languages, such as Python, Matlab, R, and Wolfram Mathematica, can be used to read it. 
    No particular dependency is required.
    
    Because the dataset contains detailed information about chemical formulas as well as the compositions of the target materials and precursors for each recipe, it can be easily used to conduct a literature review by querying desired precursors and (or) targets in different chemical spaces.
    For example, selecting all TiO\textsubscript{2} synthesized from TiCl\textsubscript{4} allows an exploration of how other synthesis formulations, such as synthesis actions, attributes, and quantities, affect the results.
    Furthermore, the materials entries in the dataset  are supplied with the Materials Project \cite{pymatgen} identifiers, thus facilitating the integration of the recipes with the thermochemical data available in the Materials Project\cite{PerssonPRB12,SunNC19}.
    
    In addition, this solution-based synthesis dataset keeps the same data structure as that in the solid-state dataset generated in our previous work \cite{SolidStateDataPaper}. 
    Therefore, it is easy to analyze the recipes from the two datasets.
    
    Despite the dataset being provided as a static snapshot, we intend to update it on a regular basis.

\section{Code availability}

The scripts used to classify paragraphs and extract recipes as well as to perform the data analysis are home-written codes which are publicly available at the github repository
\href{https://github.com/CederGroupHub/text-mined-solution-synthesis_public}{https://github.com/CederGroupHub/text-mined-solution-synthesis\_public} with acknowledgement of the current paper.

The underlying libraries used in this project are all open-source:

\emph{Tensorflow} (\href{https://www.tensorflow.org}{www.tensorflow.org})

\emph{Keras} (\href{https://keras.io}{keras.io})

\emph{SpaCy} (\href{https://spacy.io}{spacy.io}) \cite{SpaCy}

\emph{NLTK} (\href{https://www.nltk.org}{https://www.nltk.org/}) \cite{NLTK}

\emph{gensim} (\href{https://radimrehurek.com}{radimrehurek.com}) \cite{rehurek_lrec}

\emph{scikit-learn} (\href{https://scikit-learn.org}{scikit-learn.org}) \cite{scikit-learn} 

\emph{ChemDataExtractor} (\href{http://chemdataextractor.org/}{chemdataextractor.org}) \cite{ChemDataExtractor}

\emph{Material Parser} (\href{https://github.com/CederGroupHub/MaterialParser}{github.com/CederGroupHub/MaterialParser})

\emph{Borges} 
(\href{https://github.com/CederGroupHub/Borges}{github.com/CederGroupHub/Borges})

LimeSoup (\href{https://github.com/CederGroupHub/LimeSoup}{github.com/CederGroupHub/LimeSoup}).

\section{Acknowledgements}
This material is based upon work supported by the National Science Foundation under Grant No. DMR-1922372. We also thank Chris Bartel, and Amalie Trewartha for valuable discussions.

\section{Author contributions}
Z.W. developed material quantities extraction algorithm, building reaction formula algorithm, Materials Parser, operations and conditions extraction algorithms, analyzed the data, and wrote the manuscript.
O.K. developed the Materials Parser, operations and conditions extraction algorithms, analyzed the data, and wrote the manuscript.
K.C. developed operations and conditions extraction algorithms, manually inspected the dataset, and wrote the manuscript.
T.H. developed materials entity recognition algorithm, and manually inspected the dataset.
H.H. curated paragraphs databases, and developed the paragraphs classifier.
Y.F. developed building reaction formula algorithm and manually inspected the dataset.
Y.Z. analyzed the data and manually inspected the dataset.
Y.S. analyzed the data and manually inspected the dataset.
Z.C. analyzed the data and manually inspected the dataset.
W.S. analyzed the data and developed material quantities extraction algorithm.
G.C. developed approach, supervised the project, and wrote the manuscript. 
All authors discussed the results and contributed to the final manuscript.

\section{Competing interests}
The authors declare no competing interests.

\bibliographystyle{naturemag}
\bibliography{refs}

\include{tables}
\include{figures}

\end{document}

%% file: tables.tex
\linespread{1.0}
\section*{Tables}

\begin{table}[ht]
\centering
\resizebox{\textwidth}{!}{\begin{minipage}{\textwidth}
\begin{threeparttable}[b]
\begin{tabular}{|p{1.7in}|c|l|}
\hline
\rowcolor{YellowGreen}
\textbf{Data description} & \textbf{Data Key Label} & \textbf{Data Type} \\
\hline
DOI of the original paper & \texttt{doi} & \emph{string} \\
\hline
Snippet of the raw text & \texttt{paragraph\_string} & \emph{string} \\
\hline
\multirow{4}{1.6in}{Chemical formula} & \texttt{reaction} & Object (\emph{dict}): \\
 & & - \texttt{left\_side}: \emph{list} of \emph{strings} \\
 & & - \texttt{right\_side}: \emph{list} of \emph{strings} \\
\hline
Chemical formula in string format & \texttt{reaction\_string} & \emph{string} \\
\hline
\multirow{8}{1.6in}{Target material data} & \texttt{target} & Object (\emph{dict}): \\
 & & - \texttt{material\_string}: \emph{string}, \\ 
 & & - \texttt{material\_formula}: \emph{string}, \\
 & & - \texttt{composition}: \emph{list} of Objects\tnote{2}, \\
 & & - \texttt{additives}: \emph{list} of \emph{strings} \\
 & & - \texttt{elements\_vars}: \{\texttt{var}: \emph{list} of \emph{strings}\} \\
 & & - \texttt{amounts\_vars}: \{\texttt{var}: \emph{list} of Objects\tnote{3}\} \\
 & & - \texttt{oxygen\_deficiency}: \emph{boolean} \\
 & & - \texttt{mp\_id}: \emph{string} \\
\hline
List of target formulas obtained after variables substitution & \texttt{targets\_string} & \emph{list} of \emph{strings} \\
\hline
Precursor materials data & \texttt{precursors} & \emph{list} of Objects (See \texttt{target}) \\
\hline
List of solvent formulas & \texttt{solvents\_string} & \emph{list} of \emph{strings} \\
\hline
\multirow{4}{1.6in}{Sequence of synthesis steps and corresponding conditions} & \texttt{operations} & \emph{list} of Objects (\emph{dict}): \\
 & & - \texttt{token}: \emph{string}, \\
 & & - \texttt{type}: \emph{string}  \\
 & & - \texttt{conditions}: Object \\
 & & - - \texttt{temperature}: \emph{list} of Objects\tnote{4} \\
 & & - - \texttt{time}: \emph{list} of Objects\tnote{4}, \\
 & & - - \texttt{atmosphere}: \emph{list} of \emph{strings} \\
 & & - - \texttt{mixing\_device}: \emph{list} of \emph{strings} \\
 & & - - \texttt{mixing\_media}: \emph{list} of \emph{strings} \\
\hline
\multirow{4}{1.6in}{Materials with corresponding quantities} & \texttt{quantities} & \emph{list} of Objects (\emph{dict}): \\
 & & - \texttt{material}: \emph{string}, \\
 & & - \texttt{quantity}: \emph{list} of Objects\tnote{5}  \\
\hline
Synthesis type & \texttt{type} & \emph{string} \\
\hline
\end{tabular}
\begin{tablenotes}
 \item[1] \{\texttt{amount}: \emph{float}, \texttt{material}: \emph{string}\}
 \item[2] \{\texttt{formula}: \emph{string}, \texttt{elements}: \{\texttt{element}: amount of element\}, \texttt{amount}: \emph{string}\}
 \item[3] \{\texttt{max\_value}: \emph{float}, \texttt{min\_value}: \emph{float}, \texttt{values}: \emph{list} of \emph{floats}\}
 \item[4] \{\texttt{max\_value}: \emph{float}, \texttt{min\_value}: \emph{float}, \texttt{values}: \emph{list} of \emph{floats}, \texttt{units}: \emph{string} \}
  \item[5] \{\texttt{number}: \emph{float},  \texttt{unit}: \emph{string} \}
\end{tablenotes}
\end{threeparttable}
\caption{
Format of each data record: description, key label, data type. 
}
\label{tab:tab1}
\end{minipage} }
\end{table}

\begin{table}[ht]
\centering
\begin{tabular}{|l|c|c|c|c|}
\hline
\rowcolor{YellowGreen}
\textbf{Data attribute} & \textbf{Precision} & \textbf{Recall}  & \textbf{F1 score} \\
\hline
Balanced reactions & 0.94  & / & / \\
 - targets & 0.97 & / & / \\
 - precursors & 0.98 & 0.99 & 0.98 \\
\hline
Operations & 0.96 & 0.85 & 0.90 \\
\hline
Conditions & & & \\
 - temperature & 0.97 & 0.92 & 0.94 \\
 - time & 0.98 & 0.89 & 0.93 \\
 - atmosphere & 0.97 & 0.92 & 0.94 \\ 
\hline
Quantities & 0.90  & 0.85 & 0.87 \\
\hline
\end{tabular}
\caption{
Performance of data extraction for dataset entries.
}
\label{tab:tab2}
\end{table}

\begin{table}[ht]
\centering
\begin{tabular}{| c || c |}
\hline
\rowcolor{YellowGreen}
\textbf{Targets} & \textbf{Common Precursors} \\
\hline
ZnO& Zn(NO\textsubscript{3})\textsubscript{2}, Zn(Ac)\textsubscript{2}, ZnCl\textsubscript{2} \\ 
TiO\textsubscript{2} & Ti(OCH(CH\textsubscript{3}))\textsubscript{4}, Ti(OC\textsubscript{4}H\textsubscript{9})\textsubscript{4}, TiCl\textsubscript{4} \\
Fe\textsubscript{3}O\textsubscript{4}&FeCl\textsubscript{3}, FeCl\textsubscript{2} \\
Fe\textsubscript{2}O\textsubscript{3} & FeCl\textsubscript{3}, Fe(NO\textsubscript{3})\textsubscript{3}\\
SnO\textsubscript{2} & SnCl\textsubscript{4} \\
ZrO\textsubscript{2} & ZrOCl\textsubscript{2}, ZrO(NO\textsubscript{3})\textsubscript{2} \\
CuO & Cu(NO\textsubscript{3})\textsubscript{2}, Cu(Ac)2, CuCl\textsubscript{2}, CuSO\textsubscript{4} \\
SiO\textsubscript{2} & Si(OC\textsubscript{2}H\textsubscript{5})4  \\
WO\textsubscript{3} & Na\textsubscript{2}WO\textsubscript{4}, WCl\textsubscript{6}, H\textsubscript{2}WO\textsubscript{4} \\
CdS &Na\textsubscript{2}S, CH\textsubscript{4}N2S, CdCl\textsubscript{2}, Cd(NO\textsubscript{3})\textsubscript{2}   \\
\hline
\end{tabular}
\caption{
Ten most common targets in the dataset with their corresponding precursors.
}
\label{tab:tab4}
\end{table}

%% file: figures.tex
\section*{Figures}
\linespread{1.5}

\begin{figure}[h]
\centering
\includegraphics[width=\linewidth]{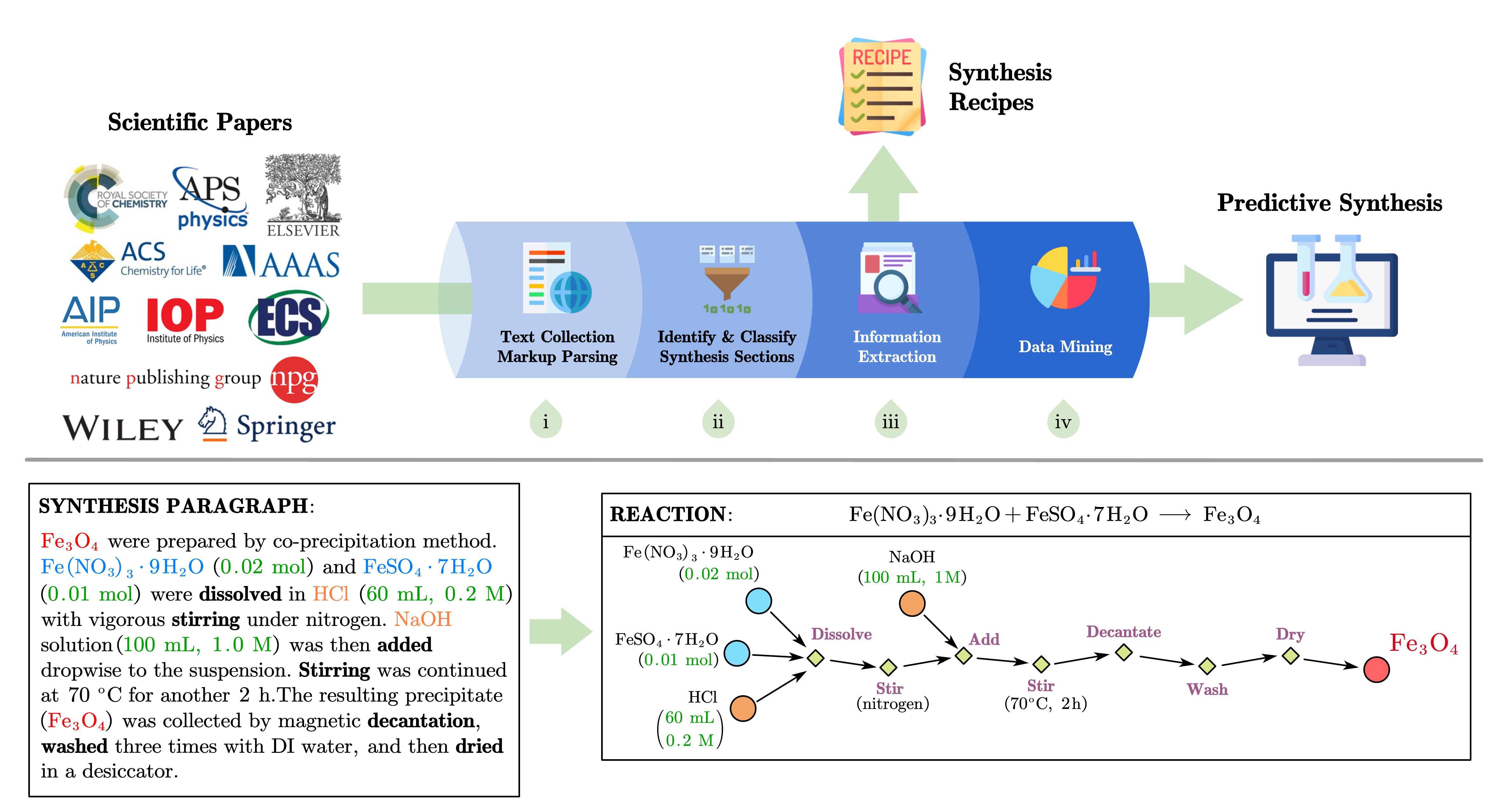}
\caption{
\textbf{Extraction pipeline and example.}
\textit{Top panel:} Schematic representation of the standard text mining pipeline: (i) scrape papers in markup format from the major publishers; (ii)  identify and classify synthesis sections; (iii) extract key information including materials, amounts, sequenced operations, and conditions; (iv) store synthesis recipes into the database for future data mining.
\textit{Bottom panel:} Example of a codified recipe extracted from a synthesis paragraph. 
}
\label{fig:fig1}
\end{figure}

\begin{figure}[ht]
\centering
\includegraphics[width=\linewidth]{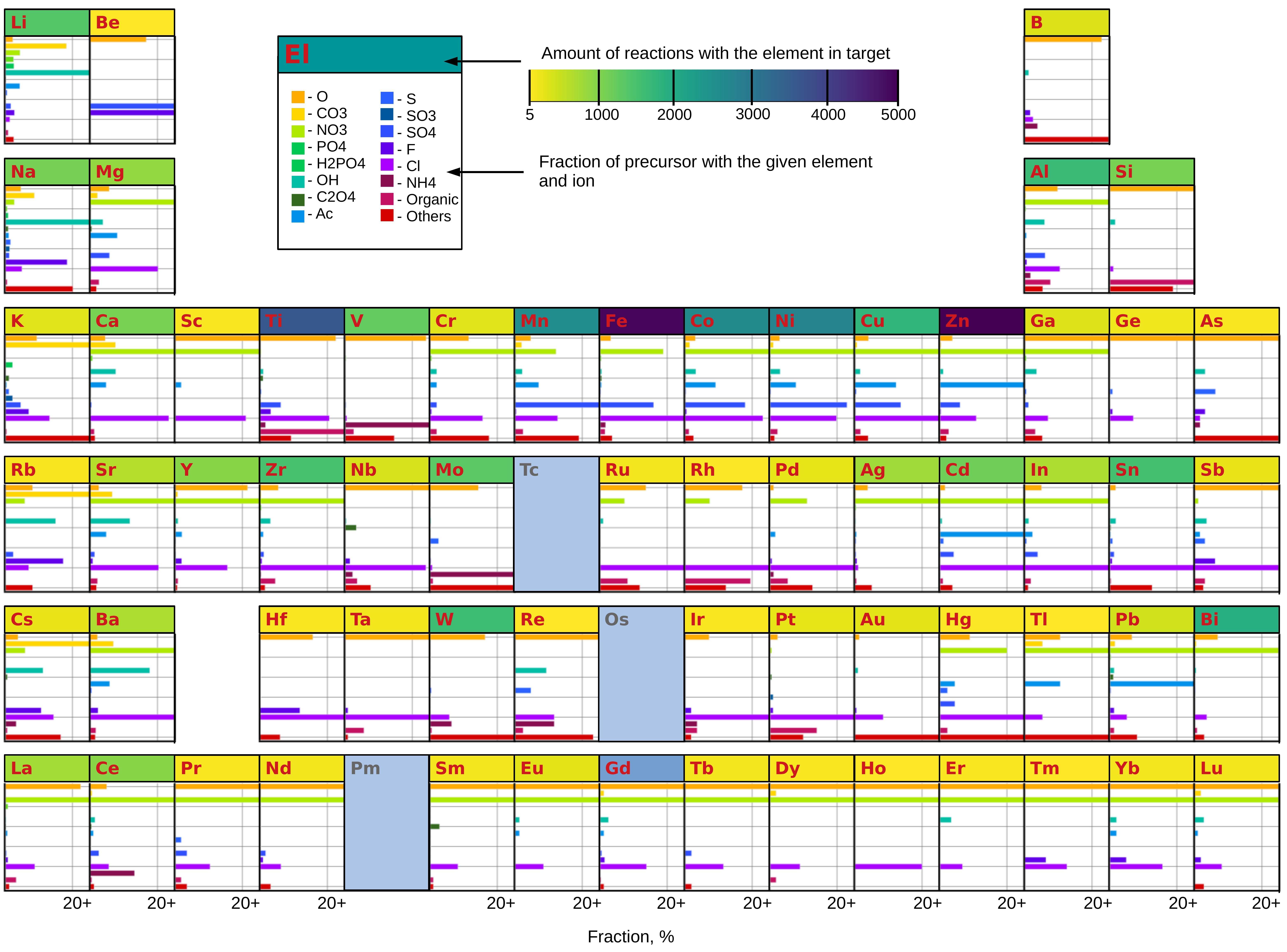}
\caption{
\textbf{The chemical space covered by the dataset.} 
For each element, the box containing the element name is colored in a yellow-to-navy blue gradient representing the total amount of reactions that produce a target compound containing the element.
The bar graph below each element shows the list of ions paired with the element in precursor compounds.
The fractions of the precursors (i.e. element+ion) used are shown by the length of the bars. 
Boxes with no bar graph represent elements occurring in five and fewer targets.
``Ac'' stands for acetate radical CH\textsubscript{3}COO\textsuperscript{--} in the compound formula.
}
\label{fig:fig2}
\end{figure}

\begin{figure}[ht]
\centering
\includegraphics[width=0.8\linewidth]{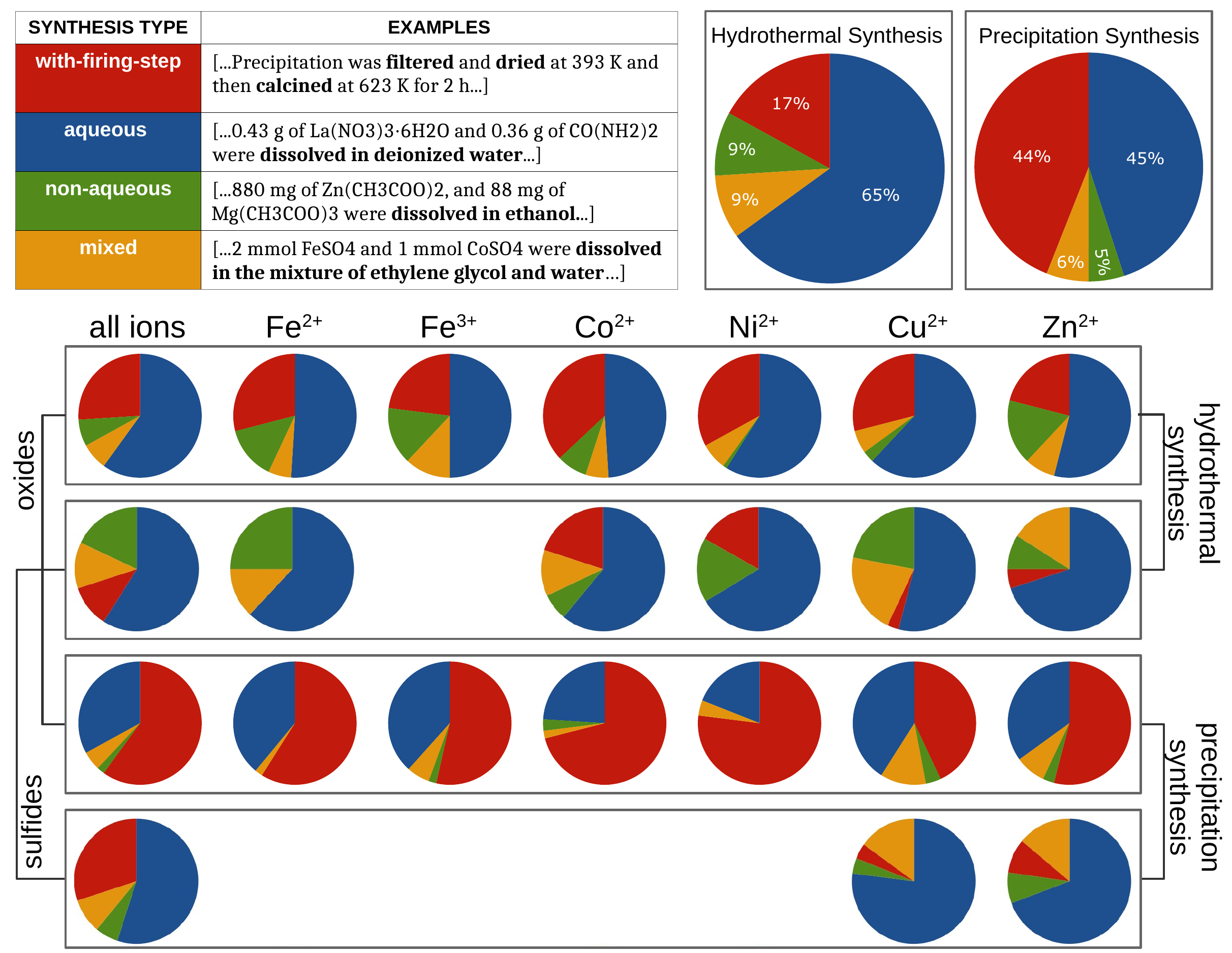}
\caption{
\textbf{Correspondence between choice of synthesis route and selected types of targets.}
The top table gives an example of the four synthesis categories defined: with firing step, aqueous, non-aqueous, and mixed.
The two pie-charts on the top-right show the fractions of synthesis routes in the hydrothermal and precipitation datasets separately.
The four rows of pie charts in the lower half of the figure represent the fractions of the four synthesis routes (given in the table) for all oxides, all sulfides, and individual oxides and sulfides with different oxidation states of data-rich transition metals separately.
The first and second rows are results from the hydrothermal dataset.
The third and fourth rows are results from the precipitation dataset.
Each blank space means that there is not enough data to form a statistic for the corresponding type of target.
}
\label{fig:fig3}
\end{figure}